\documentclass[a4]{scrartcl}

\usepackage[utf8x]{inputenc}
\usepackage{amsmath,amssymb}
\usepackage{bm} 
\usepackage{graphicx}
\usepackage{color}
\usepackage{dcolumn}
\usepackage{bm}
\usepackage[numbers,super,comma,sort&compress]{natbib}
\usepackage{units}
\usepackage[super]{nth}
\usepackage{algorithm}
\usepackage{algpseudocode}
\usepackage{siunitx}
\usepackage[nomessages]{fp}
\usepackage{hyperref}
\usepackage{booktabs}
\usepackage{multirow}
\usepackage{xspace}

\definecolor{background-color}{gray}{0.98}
\definecolor{todo}{rgb}{1,0,0}
\title{Compressing molecular dynamics trajectories: breaking the one-bit-per-sample barrier}
\author{
  Jan Huwald\thanks{Biosystem Analysis Research Group \newline Department of Mathematics and Computer Science and Jena Centre for Bioinformatics \newline Friedrich Schiller University Jena \newline Ernst-Abbe-Platz 1-3 / 07743 Jena / Germany \newline Correspondence: jh@sotun.de / peter.dittrich@uni-jena.de},
  Stephan Richter\footnotemark[1],
  Peter Dittrich\footnotemark[1]
}

\newcommand{\huri}{\textsc{Hrtc}\xspace}

\begin{document}

\maketitle

\begin{abstract}
Molecular dynamics simulations yield large amounts of trajectory data.
For their durable storage and accessibility an efficient compression algorithm is paramount.
State of the art domain-specific algorithms combine quantization, \textsc{Huffman} encoding and occasionally domain knowledge.

We propose the high resolution trajectory compression scheme (\huri)
  that relies on piecewise linear functions to approximate quantized trajectories.
By splitting the error budget between quantization and approximation,
  our approach beats the current state of the art by several orders of magnitude given the same error tolerance.
It allows storing samples at far less than one bit per sample.
It is simple and fast enough to be integrated into the inner simulation loop,
  store every time step,
  and become the primary representation of trajectory data.
\end{abstract}

\bibliographystyle{acm}

\section*{Introduction}

Molecular dynamic (MD) simulations are among the largest supercomputer uses.
Computing power increases exponentially faster than communication bandwidth.\cite{Esmaeilzadeh2011}
To retain the ability to durably store, share and even analyze the generated particle trajectories,
  they have to be represented efficiently.

For example, a recent atomistic model of the SGLT membrane protein,
  consisting of 90.000 particles simulated for $2.4 \times 10^8$ steps ($\SI{480}{ns}$)
  generates $\SI{259}{TiB}$ of raw trajectory data.\cite{Adelman2014}
The \textit{de-facto} standard approach to handle such large datasets at all,
  is to down-sample the time-domain of the trajectory to a tiny fraction---in
  said example by saving only \nicefrac{1}{50000} of the steps,
  except for a few spotlight situations where \nicefrac{1}{500} of all timesteps are saved.

Typically, the down-sampled trajectories are then further compressed.
In principle, this is possible using a general purpose lossless compression algorithm, e.g. \textsc{BZip2}.
Unfortunately, general purpose compression suffers from incompressible noise in the less significant bits of the particle positions.
They can at most be considered as base line to compare better algorithms against.

MD trajectories are highly amendable to special-purpose compression:
The interframe variation of the particle positions is orders of magnitudes smaller than the positions themself ($\Delta x \ll x$).
The demanded precision is typically much smaller than the precision offered by uncompressed representation ($\SI{32}{b}$ or $\SI{64}{b}$ in IEEE 754).
Positions as well as velocities are strongly correlated with values from the past and neighboring particles.

Furthermore, MD trajectory compression has special requirements not fulfilled by general-purpose algorithms.
For details, see \textsc{Marais} et al.\cite{Marais2012} and especially \textsc{Spångberg} et al.\cite{Spangberg2011}
We concentrate on three aspects:
\begin{itemize}
\item \emph{Speed}: 
  (de-)compression overhead has to be insignificant compared to the simulation itself to be of any use.
  The simulator itself is a space-optimal compressor, requiring only the initial state and the elapsed time to be stored.
  Corollary it has to be
    \emph{parallelizable},
    computable in a \emph{streaming} fashion,
    have a \emph{small memory footprint}, and
    elide random writes to the underlying storage.
\item \emph{Tunability}:
  The tolerable error of lossy compression is highly dependent on the simulated scenario and intended analysis.
  A tunable precision and good performance across all values is thus required.
\item \emph{Simplicity}:
  Complex code fits neither into processor caches nor into programmer minds.
  It is thus prone to be slow, faulty and not widely implemented.
\end{itemize}

With the advent of large datasets, a number of compression schemes have been proposed.
Most of them are a combination of the following building blocks:
\begin{itemize}
\item Quantization:
  the lossy reduction to represent floating point numbers as small integers ($x \mapsto \lfloor 2 x / \epsilon_q +0.5 \rfloor$).
  The quantization error $\epsilon_q$ is a paramount tunable of these algorithms.
\item Delta-coding:
  storing the difference of consecutive values instead of the values itself
    ($x_0, x_1, x_2, \dots \mapsto x_0, x_1 - x_0, x_2 - x_1, \dotsc$).
\item Reordering
  particle coordinates, so that their consecutive differences are likely to be small.
  This comes often at a loss of the particle identities:
    particles of the same element become indistinguishable.
\item Variable-length integer encoding
  to store the frequent small integers with few bits without sacrificing the possibility to store rare large values.
  This is typically achieved using
    \nth{0} order encoders, which are faster but rely on a fixed distribution of values (e.g., rice coding), or
    \nth{1} order encoders, which adapt to the observed distribution to increase the compression rate
      at the price of higher space and time complexity (e.g., \textsc{Huffman} coding).
\end{itemize}
All of these techniques have in common that they try to spend less bits per data-point,
  but keep the number of data-points constant.
They thus fail to achieve less than one bit per sample.

\textsc{Spångberg} et al.~proposes \textsc{Tng-Mf1}\cite{Spangberg2011}, a class of algorithms that use
  quantization,
  delta-coding within and between frames,
  a custom \nth{0} order variable length integer compression, and
  optionally a combination of \textsc{Burrow-Wheeler} transformation\cite{Burrows1994},
    \textsc{Lempel-Ziv} coding\cite{Lempel1977}, and \textsc{Huffman} coding.\cite{Huffman1952}
\textsc{Marais} et al.\cite{Marais2012}~use
  quantization,
  an arithmetic encoder, and
  interframe prediction with polynomials of order zero or one.
Additionally they use \textit{a~priori} knowledge about the spatial structure of water
  to exploit redundancy in adjacent water molecules position and orientation.
The venerable \textsc{Xtc} file format uses
  quantization,
  delta-coding between frames,
  reordering of coordinates to improve compression of water molecules, and
  a custom variable-length integer encoding.\cite{Abraham2015}

A completely different approach is realized with the Essential Dynamics tool,
which stores particle motions relative to a reference structure in a matrix.
The matrix' eigenvectors with the largest eigenvalues are used as a compressed base:
a weighted sum of them represents each frame.\cite{Meyer2006}
\textsc{Ohtani} et al.~represent trajectories by polynomial functions.\cite{Ohtani2013}
A time window for consecutive frames is decreased until a polynomial function fits the data within the given error.
Both methods do not support streaming operation and have large time and space overhead.
For more examples, see \textsc{Marais et al.}\cite{Marais2012} 

A related field to MD trajectory compression is the efficient storage of space curves used for geoinformation systems.
These algorithms do not store time information, work offline on the entire dataset, and are allowed superlinear runtimes.
The prime example is the \textsc{Douglas-Peucker} algorithm that iteratively removes points from a curve,
  as long as they lie within an error corridor between their neighbors.\cite{douglas1973algorithms}
\textsc{Bellman}'s algorithm even finds the optimal (minimal error) cover of $n$~points with $k$~lines,
  but requires $O(n^2)$ time to do so.\cite{bellman1961approximation}

The high resolution trajectory compression algorithm (\huri) presented here follows a similar, yet faster and simpler approach:
Akin to delta-coding, piecewise linear functions are employed to represent trajectories.
The resulting support vectors are
  quantized with a tunable precision and
  stored using state-of-the-art variable length integer representation.
Besides having the highest compression rates and performance,
our main novelty is the distribution of the error budget between the quantization and the approximation by functions.
Established approaches allow either quantization error\cite{Marais2012,Spangberg2011,Abraham2015},
  or approximation error\cite{Ohtani2013}, but not both at the same time.

Comparing our algorithm to the state of the art is difficult:
Crucial simulation parameters are not well documented and
  a common standard for storing trajectory data needs to be established.\cite{Hinsen2014}
For this purpose we heavily rely on the container-based file format
  proposed by \textsc{Lundborg} et al.:\cite{Lundborg2014}
It allows trajectory storage alongside with parameter values and arbitrary metadata in a single file.
It enables usage of different compression algorithms,
initially equipped with the option to use \textsc{Xtc}, \textsc{Tng} and \textsc{BZip2}.
Both the file format and the code are open source and open to extension.
We followed this invitation and modified the \textsc{Tng} library to support our \huri compression scheme.

\section*{Methodology}

The \huri algorithm we present here uses piecewise linear functions to represent trajectories.
We consider only particle positions---noting that for most applications the slope of the approximating function can be used as velocity.
Given $n$ particles in a $d$-dimensional space, we consider each dimension $k \in \{1, \dotsc, n d\}$ of a state independently.
For each dimension $k$, our algorithm approximates the
  $T$ points $(x_{k,t})$ of a trajectory by
  $J(k)$ functions $f_{k,j} : \{1, \dotsc, \Delta t_{k,j}\} \rightarrow \{\frac{1}{2} z \epsilon_q : z \in \mathbb{Z}\}$ where
  $\epsilon_q$ is the quantization error, and
  $\Delta t_{k,j}$ is the number of timesteps the function $f_{k,j}$ is covering, with
  $j \in \{ 1, \dotsc, J\}$,
  $t \in \{ 1, \dotsc, T \}$.
With a given (maximal) approximation error $\epsilon_f$, our approximation scheme
becomes:
\begin{align}
x_{k,1}, \dotsc, x_{k,T} \rightsquigarrow f_{k,1}, \dotsc, f_{k,J(k)} ~,\\
|x_{k,t} - f_{k,j}(t - \sum_{l<j} \Delta t_{k,l})| \le \epsilon_f ~,\\
\sum_{l<j} \Delta t_{k,l} < t  \le \sum_{l<j+1} \Delta t_{k,l} ~.
\end{align}
A high compression rate can be achieved by covering large durations (maximize $\Delta t_{i,j}$) with functions that require few bits to encode.
In contrast to \textsc{Tng}, \textsc{Xtc} and other classical algorithms 
  we introduce multiple points of information loss during compression.
That is, we split the total error budget $\epsilon$ into two parts:
  the inevitable quantization error $\epsilon_q$, and
  the approximation error $\epsilon_f$.
For this work we use by default $\epsilon_q = \epsilon_f = \frac{1}{2} \epsilon$.

Exploiting the approximation error $ \epsilon_f$ is our primary vehicle for high compression rates:
With $\epsilon_f = 0$ a polynomial representing $n$ points requires $n$ support vectors.
Storing them would be no more efficient than storing the points themselves. 
By introducing an error $\epsilon_f > 0$, multiple functions are valid representations of the point sequence.
From this set of functions, we can select those with the fewest support vectors.

In \huri, we implicitly maintain a set of functions that are valid approximations for each dimension.
Every time a new point is added to the sequence,
we remove those functions from the set that are not valid approximations of the extended sequence.
We continue this process until just before adding the next point would render the candidate set empty.
This implements an abstract 
greedy search for valid approximations covering the maximal timespan $\Delta t_{i,j}$ (see Algorithm~\ref{alg:prelim}).

\begin{algorithm}
\caption{Abstract algorithm using arbitrary functions for approximation}
\begin{algorithmic}[1]
\State $F \leftarrow$ set of all approximation functions
\State $\Delta t \leftarrow 0$  

\For{$t \in \{0,\dotsc,T\}$}
  \State $p' \leftarrow  p~\text{appended by} \lfloor  x_t / (2 \epsilon_q) +0.5 \rfloor$ 
      \Comment Quantize input 
  \State $F' \leftarrow$ valid approximation functions for $p'$

  \If{$F' = \emptyset$ \text{or} $t = T$}
    \State output $(F, \Delta t)$
    \State $p \leftarrow (x_t)$
    \State $F \leftarrow$ set of all functions
    \State $\Delta t \leftarrow 0$
  \Else
    \State $p \leftarrow p'$
    \State $F \leftarrow F'$
    \State $\Delta t \leftarrow \Delta t + 1$
  \EndIf
\EndFor
\end{algorithmic}
\label{alg:prelim}
\end{algorithm}

For general functions this algorithm is expensive in time and space:
Even if restricted to polynomials with integer coefficients, explicitly storing all functions has exponential space complexity.
Generating the new function candidate set $F'$ in line 5 requires at least
  looking at all $\Delta t$ values, lifting the lower time bound to $\Omega(n\cdot \Delta t)$.

To remain in our time budget we restrict the function space to linear functions.
The first support vector is the final point of the previous interval,
  the second one at the rightmost point of the current interval. 
Linear functions allow us to store the entire candidate set $F$ using two integers,
  and merging as well as computing the candidate set in constant time.

\begin{figure}
\def\svgwidth{0.3\linewidth}\input{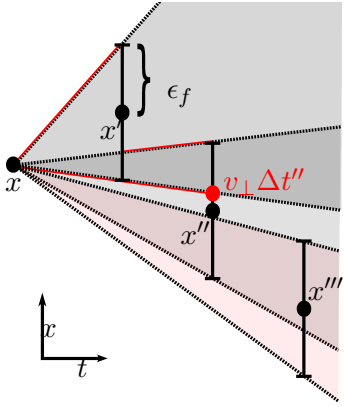}
\caption{\label{fig:illustrated_segment}
  Illustration of a segment used to capture a part of the trajectory:
  The leftmost node $x$ is the left support vector of the current one.
  Together with the error bars around each node $x' \ldots x'''$, it induces the extremal slopes $v_{\bot,i}, v_{\top,i}$ (dotted lines).
  The shaded cones cover all possible slopes for each point.
  While successively including more points into the segment, the range of valid slopes $[v_\bot, v_\top]$ (red line) decreases.
  The error cone induced by $x'''$ (shaded red) does not intersect with  $[v_\bot, v_\top]$ covering $x$, $x'$ and $x''$ (shaded dark-grey).
  Thus $v_\bot \Delta t''$ becomes the terminal node of this segment: it is the valid point closest to $x''$.
}
\end{figure}

The set of linear functions through a point $x$ that are valid approximations
  for a second point $x'$ separated by $\Delta t$ timesteps
  form a $2 \epsilon_f$~wide error cone around $x'$ (see Figure~\ref{fig:illustrated_segment}).
The set is denoted by $\mathrm{AF}$:
\begin{align}
  & \{ f : f(0) = x \wedge | f(\Delta t) - x' | \le \epsilon_f  \} \nonumber \\
  \supseteq & \{ t \mapsto \frac{t}{\Delta t} v  + x : v \in [v_\bot, v_\top] \} \\
  =: & \mathrm{AF}(v_\bot, v_\top) \quad \text{with} ~
    v_\bot = \frac{x' - x - \epsilon_f}{\Delta t},\;
    v_\top = \frac{x' - x + \epsilon_f}{\Delta t} ~.
\nonumber
\end{align}
The set of functions is completely represented by $x$, $v_\bot$, and $v_\top$.
The intersection of multiple such sets
  that share a common support vector $x$ but differ in extremal slopes $v_{\bot,i}$ and $v_{\top,i}$
  can be merged efficiently:
\begin{align}
\bigcap_{i} AF(x, v_{\bot,i}, v_{\top,i}) = AF(x, \max_{i} \, v_{\bot,i} , \min_{i} \, v_{\top,i}) ~.
\end{align}
This intersection contains all valid approximations for \emph{all} input points.
Note that the intersection is empty if the lower bound $v_\bot$ becomes larger
than the upper bound $ v_\top$, that is,  $AF(x, v_\bot, v_\top) = \emptyset$ iff.~$v_\bot > v_\top$.
The above formulation allows to incrementally update the function candidate set.
The sequence of points of the current interval does not have to be stored anymore.

\subsection*{Combined storage of multiple trajectories}

A single trajectory can be stored 
in memory as a sequence $(v_0, \Delta t_0), (v_1, \Delta t_1), \dotsc$
When storing multiple trajectories, we have to counter the issue that their support vectors are not synchronized in time.
A naive approach would include a trajectory index $k_l$ for each support vector:
  $(k_1, v_{k_1}, \Delta t_{k_1}), (k_2, v_{k_2}, \Delta t_{k_2}), \dotsc$.
However, this would require $O(\log n d)$ additional space per support vector, which we avoid with the following procedure.

We exploit the fact that the start time and duration of a segment imply the time at which the next support vector has to be expected.
To use this insight, we maintain an auxiliary priority queue $Q_\mathrm{expected}$, which
stores tuples $(t, k)$ sorted by time $t$ and secondarily trajectory index $k$.
The minimal element of the queue denotes the support vector to be outputted next.
The queue always contains $nd$ elements,
  thus the space gain is traded against an additional time complexity of $O(\log nd)$ per inserted interval.

To reorder the support vectors from the sequence of discovery to the sequence of storage described above,
  we use a second priority queue $Q_\mathrm{known}$.
It stores all support vectors that can not be stored immediately because a preceding support vector (according to $Q_\mathrm{expected}$) is not yet known.
It contains up to $O(nd \max_i \Delta t_i)$ elements.
This implies a memory overhead of the same magnitude in the compressor.
The time overhead can be reduced to $O(nd)$ assuming time points are dense--if
  the number of support vectors of all trajectories exceeds the number of time points.
Due to the blockwise compression and limited $\Delta t$ observed,
  this overhead is much less dramatic in practice.

A key frame is used to initialize the queue and to provide the initial support vectors for each dimension.
The key frame stores the quantized values of $x_{0,0}, x_{1,0}, \dotsc$ with a fixed bit count, without further compression.
The number of bits $\lceil log \frac{L}{2 \epsilon_q} \rceil$ is determined by
  the quantization granularity $\epsilon_q$ and
  the edge length $L$ of the bounding box of the trajectory. 

\subsection*{Hierarchy of computation and storage}

\begin{figure}
\includegraphics[width=\linewidth]{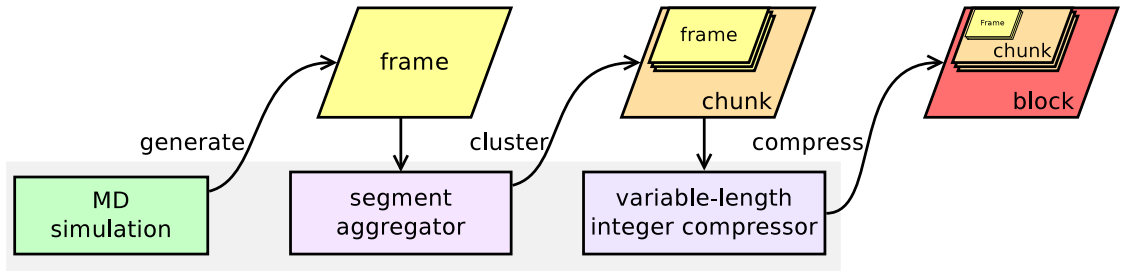}
\caption{\label{fig:storage_hierarchy}
  Hierarchical storage of MD simulation generated data.
  Top row denotes data, bottom row code. Each block starts with a key frame.
  A high-performance implementation typically integrates all lower boxes.}
\end{figure}

The \huri algorithm has to conciliate between two opposing forces:
Memory pressure and data cache locality demand to minimize the window of trajectory data held in memory during compression.
Higher compression rates require a larger timespan of data to reason about.
To minimize cache thrashing, switching between simulation and compression should happen at a low frequency---asking
  for a large window, too.
At last, to generate a searchable compressed data-stream, key frames have to be inserted at regular time intervals.

Balancing these demands has led to the following hierarchy of computation and storage (cf.~Figure~\ref{fig:storage_hierarchy}):
An MD simulator calls the compression library for each generated timestep---denoted a \emph{frame}.
The data is approximated with linear functions, their resulting support vectors are buffered.
Once a threshold buffer size is reached (e.g., $\SI{8}{KB}$ storing $1024$ support vectors),
  all support vectors are fed through the variable length integer encoding,
  yielding one \emph{chunk}.
The largest unit---a \emph{block}---contains a user-specified number of frames.
It starts with a single key frame and contains all chunks belonging to the encoded frames.
The block size is tuned by the user depending on the desired compression and seek time.

The \huri compression integrated into the \textsc{Tng} library does not fully exploit this hierarchy.
Due to design constraints of the \textsc{Tng} library,
  all frames of a block are collected and then compressed at once.
When speed matters, \huri could be used directly.

The update compression algorithm, together with storage hierarchy and multiple trajectory storage
are described by Algorithm~\ref{alg:alg}. The respective decompression is described by Algorithm~\ref{alg:sffam-large-hier}.

\begin{figure}
\def\svgwidth{\linewidth}\input{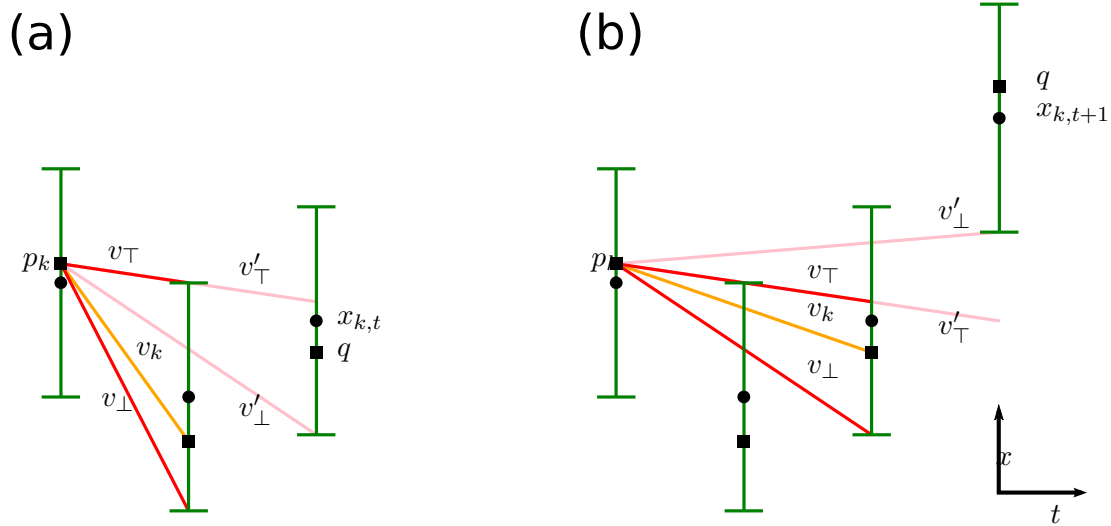}
\caption{\label{fig:pseudocode_segement}
  Illustration of the variables used during compression for each dimension $k$ as used in algorithm~\ref{alg:alg}.
  Shown are two cases: A point to be added either extends the segment (a),
    or causes termination of the segment at the previous point (b).
  The input $x_{k,t}$ is shown as circle, the quantized values as square.
  Four variables are stored per dimension:
    $p_k$ denotes the quantized support vector starting the segment,
    $v_k$ the slope to the last point added, and
    $v_\bot$ and $v_\top$ the lower and upper slope bounds.
  In addition, three temporary variables are used:
    the quantized input value $q$, and
    the updated slope bounds $v_\bot'$ and $v_\top'$.}
\end{figure}

\renewcommand{\baselinestretch}{1}
\begin{algorithm}
\caption{The \huri compression algorithm. For an illustration of the variables used, see Figure~\ref{fig:pseudocode_segement}.}
\begin{algorithmic}[1]
\State \textbf{Input:} Trajectory $(x_{t,k})$ for $t \in \{0, \dots, T\}, k \in \{1, \dots, nd \}$, quantization error $\epsilon_q$, approximation error $\epsilon_f$
\State \textbf{Output:} Compressed trajectory (incl. key frames)
\State \textbf{global} $v_{\bot,k}$, $v_{\top,k}$, $v_k$, $p_k$, $\Delta t_k$ for $k \in \{1, \dots, nd \}$ \Comment current approximation
\State \textbf{global} $q$ \Comment current quantized input $x_{t,k}$  
\State \textbf{global} outputBuf $= ()$
\State \textbf{global} $Q_\mathrm{known} \leftarrow \emptyset$,
       $Q_\mathrm{expected} \leftarrow \emptyset$
\Statex
\For{$t \in \{0, \dotsc, T\}$}                              \Comment For each input frame
  \State \Call{DrainQueue}{}                                \Comment Move expected known segments to output buffer

  \If{$t ~\mathrm{mod}~ \mathrm{blockSize} = 0$}
    \Comment ENCODE KEY FRAME
    \While{$Q_\mathrm{expected} \ne \emptyset$}              \Comment While there are segments to end
      \State $(t', k) \leftarrow \min(Q_\mathrm{expected})$ \Comment Get time and dimension of next segment
      \State \Call{FlushSegment}{t', k}                    \Comment End this segment and push it into $Q_\mathrm{known}$ 
    \EndWhile
    \State \Call{DrainQueue}{} \Comment Move $Q_\mathrm{known}$ (known segments) to output buffer
    \State \Call{FlushChunk}{}                             \Comment Compress output buffer and output it 
    \State $Q_\mathrm{expected} \leftarrow \{(t,0), \dotsc, (t, nd-1)\}$   \Comment Initialize  $Q_\mathrm{expected}$ with current time 
    \For{$k \in \{1, \dotsc nd\}$}                                        \Comment For each dimension $k$:
      \State $p_k \leftarrow \lfloor  x_{t,k} / (2 \epsilon_q) + 0.5 \rfloor$  \Comment get the next input, quantize and remember it as $p_k$,
      \State output $p_k$                                                  \Comment and output it (for the key frame).
      \State $v_{\bot,k} = -\infty$, $v_{\top,k} = \infty$    \Comment initialize empty segment
      \State $\Delta t_k \leftarrow 0$                           
    \EndFor
  \Else \Comment ENCODE SEGMENTS 
    \For{$k \in \{1, \dotsc nd\}$} \Comment for each dimension $k$
      \State $q \leftarrow \lfloor  x_{t,k} / (2 \epsilon_q) + 0.5 \rfloor$ \Comment quantize next input 
      \State $v_\top' \leftarrow \min(v_{\top,k}, (q -  p_k + \epsilon_f)/\Delta t_k)$ \Comment Computer lower and upper bound 
      \State $v_\bot' \leftarrow \max(v_{\bot,k}, (q -  p_k - \epsilon_f)/\Delta t_k)$ \Comment of the approximating functions.
      \If{$v_\top' < v_\bot'$}   \Comment If no valid approximating function remains
       \State \Call{FlushSegment}{t, k}  \Comment terminate current segment with previous point.
      \Else
        \State $v_{\bot,k} \leftarrow v_\bot'$, $v_{\top,k} \leftarrow v_\top'$ \Comment Update current approximation
        \State $v_k \leftarrow (q -  p_k) / \Delta t_k$ 
        \State $\Delta t_k \leftarrow \Delta t_k + 1$
      \EndIf
    \EndFor
  \EndIf
\EndFor
\algstore{algcomp}
\end{algorithmic}
\label{alg:alg}
\end{algorithm}

\begin{algorithm}
\begin{algorithmic}[1]
\algrestore{algcomp}
\Function{FlushSegment}{t, k}
  \State \(\Delta q \leftarrow \left\{ \begin{array}{ll}
    \Delta t_k v_{\bot,k} & \mathrm{if}\: v_k < v_{\bot,k} \\
    \Delta t_k v_k & \mathrm{if}\: v_k \in [v_{\bot,k}, v_{\top,k}] \\
    \Delta t_k v_{\top,k} & \mathrm{if}\: v_k > v_{\top,k} \end{array}\right.\)  \Comment Choose support vector closest to prev. point.
  \State insert $(t - \Delta t_k, k, \Delta t_k, \Delta q)$ into $Q_\mathrm{known}$ \Comment Insert support vector to $Q_\mathrm{known}$, 
  \Statex \Comment which is sorted by $(t - \Delta t_k, k)$.
  \State $\Delta t_k \leftarrow 1$  \Comment Initialize next segment duration with 1
  \State $v_{\bot,k} \leftarrow q - p_k - \epsilon_f$ \Comment Reset upper and lower bounds
  \State $v_{\top,k} \leftarrow q - p_k + \epsilon_f$
  \State $p_k \leftarrow p_k + \Delta q$  \Comment Set starting point of next segment 
  \Statex \Comment to terminal point of current segment.
\EndFunction
\Statex
\Function{FlushChunk}{}
  \State compress outputBuf using \textsc{VSE-R}, output result
  \State outputBuf $\leftarrow ()$
\EndFunction
\Statex
\Function{DrainQueue}{$Q_\mathrm{known}, Q_\mathrm{expected}, \mathrm{outputBuf}$}
  \While {$\min(Q_\mathrm{expected}) \overset{t,k}{=} \min(Q_\mathrm{known})$}
    \Comment While next segment is known
    \State $(t, k, \Delta t, \Delta q) \leftarrow \mathrm{extractMin}(Q_\mathrm{known})$ \Comment Get next known segment
    \State $\mathrm{extractMin}(Q_\mathrm{expected})$ \Comment Update the expected start time 
    \State insert $(t+\Delta t, k)$ into $Q_\mathrm{expected}$ 
    \State $\mathrm{outputBuf} \leftarrow (\Delta q, \mathrm{outputBuf}, \Delta t)$ \Comment Append segment to output buffer 
 \Statex \Comment (see section ``chunk encoding'')
    \If{length of outputBuf $\ge 2 \cdot \mathrm{chunkLength}$}  \Comment If chunk is full.
      \State \Call{FlushChunk}{} \Comment compress and output chunk.
    \EndIf
  \EndWhile
\EndFunction
\end{algorithmic}
\end{algorithm}

\begin{algorithm}
\caption{The \huri \textbf{de}compression algorithm.
  The handling of chunks for integer compression has been omitted for brevity---it is implied in line 11.}
\label{alg:sffam-large-hier}
\begin{algorithmic}[1]
\State \textbf{Input:} Compressed trajectory (incl. key frames)
\State \textbf{Output:} Uncompressed trajectory $(x_{t,k})$
\For{$t \in \{0, \ldots, T\}$}
  \If{$t ~\mathrm{mod}~ \mathrm{blockSize} = 0$} \Comment decode key frame
    \For{$k \in \{1, \dotsc nd\}$}
      \State read $q$
      \State $p_k \leftarrow q \epsilon_q$, $t_k \leftarrow t, v_k \leftarrow 0 ,\Delta t_k \leftarrow 0$ 
                                                    \Comment Set starting point of segment
    \EndFor
    \State $Q_\mathrm{expected} \leftarrow \{(t,1), \dotsc, (t, nd)\}$ \Comment Initialize $Q_\mathrm{expected}$ with current time
  \EndIf
  \While{$t \overset{t}{=} \min(Q_\mathrm{expected})$} \Comment For all segments starting at $t$
    \State $(t, k) \leftarrow \mathrm{extractMin}(Q_\mathrm{expected})$ \Comment Get dimension of next segment
    \State read $(d, q)$ using \textsc{VSE-R} decompression \Comment Get its duration and support vector
    \State $p_k \leftarrow p_k + \Delta t_k v_k$ \Comment Compute starting point
    \State $v_k \leftarrow q \epsilon_q/d$ \Comment Compute slope of segment
    \State $t_k \leftarrow t$ \Comment Update time of segment
    \State $\Delta t_k \leftarrow d$ \Comment Duration of segment 
    \State insert $(t + d, k)$ into $Q_\mathrm{expected}$ \Comment Update expected time for next segment
  \EndWhile
  \For{$k \in \{1, \dotsc nd\}$} \Comment Interpolate current frame
    \State output $x_{t,k} \leftarrow p_k + (t - t_k) v_k$ \Comment and output it. 
  \EndFor
\EndFor
\end{algorithmic}
\end{algorithm}
\renewcommand{\baselinestretch}{1.5}

\subsection*{Integer compression and chunk encoding}

For variable-length integer encoding, we use the \textsc{Integer Encoding Library}.\cite{integer-encoding}
It offers several codecs.
After selecting for time and speed, we chose the codec \textsc{VSE-R}.\cite{silvestri2010vsencoding}
The library only encodes unsigned integers.
Where signed integers occur in our algorithm they are mapped to unsigneds:
\begin{equation}
i \mapsto \left\{ \begin{array}{lr}
  \hphantom{-} 2i   & \textrm{if } i \ge 0 ~,\\
              -2i+1 & \textrm{if } i <   0 ~.
\end{array} \right.
\end{equation}

\textsc{VSE-R} encodes groups of consecutive integers with the number of bits required by the largest element of the group.
The optimal length of the group is computed using dynamic programming.
This allows storing the number of bits only once for several integers to be stored.
It is the basis for high performance of \textsc{VSE-R} (regarding throughput and compression).
It also means that the performance is suboptimal when encoding integers of alternating magnitude.
In our case the magnitude of space and time deltas can be different by several orders of magnitude. 
So we rearrange the support vectors such that time and space deltas are grouped together, respectively.
Instead of the queue $(\Delta x_1, \Delta t_1), \ldots, (\Delta x_n, \Delta t_n)$
   we store the doubled ended queue $\Delta x_n, \ldots \Delta x_1, \Delta t_1, \ldots, \Delta t_n$ 
  (see function $\textsc{DrainQueue}$ in Algorithm~\ref{alg:alg}).

\subsection*{Optional adaptation for deep simulator integration}

Because the computational demands for our compression method are small, 
we envisage future integration of it into the inner loop of MD simulation programs:
Every update of a particle's state is immediately followed by the compression of the new position.
\huri can then serve as the primary mechanism to retrieve simulation data. In the following, we
describe optional adaptations of our algorithm for this purpose, which are, however, not
applied for our performance evaluation in the result section.



For a state-of-the-art MD application the integration of \huri in its inner loop poses additional challenges.
High performance MD simulators rely on specialized hardware---from GPUs to custom ASICs.\cite{Shaw2007}
The performance characteristic of these platforms differs from a typical desktop CPU.
Non-uniform memory access, diverging control flow and branches are much more expensive compared to arithmetic operations.
Addition and multiplication are especially fast compared to division and other mathematical operations.\cite{cuda}
To accommodate \huri compression on these machines, we can adapt the algorithm.
The critical section of our algorithm is the check
  whether the current candidate set of curves is empty after adding the next point.
We reformulate it to
  avoid division operations, and
  rely on conditional writes instead of branches.
Then the special purpose hardware only needs to transfer id and position of those dimensions where the check failed.
The host computer executes all further compression steps (sorting, queue management, variable-length integer encoding)
  in parallel to the kernel running on special purpose hardware.

The original condition whether a point does not lie in the current set of linear functions is:
\begin{align}
v_\bot' & > v_\top' \label{eqn:test_orig} ~,\\
v_\bot' & = \max(v_\bot, \frac{x' - x - \epsilon_f}{\Delta t}) \label{eqn:test_min} ~,\\
v_\top' & = \min(v_\top, \frac{x' - x + \epsilon_f}{\Delta t}) \label{eqn:test_max} ~.
\end{align}

This code requires $4$ additions, $2$ divisions, $3$ comparisons and $3$ branches.
Equation \ref{eqn:test_orig}-\ref{eqn:test_max} can be merged and simplified by case analysis:
\begin{align}
v_\bot' > v_\top' \label{eqn:test_merge}
\Leftrightarrow
  \left\{ \begin{array}{lllr}
  v_\bot                  & > \frac{x'-x+\epsilon_f}{\Delta{t}} & \text{if } v_\bot \ge \frac{x'-x-\epsilon_f}{\Delta{t}} \wedge v_\top \ge \frac{x'-x+\epsilon_f}{\Delta{t}} ~,& (\ref{eqn:test_merge}.1) \\
  v_\bot                  & > v_\top                  & \text{if } v_\bot \ge \frac{x'-x-\epsilon_f}{\Delta{t}} \wedge v_\top <   \frac{x'-x+\epsilon_f}{\Delta{t}} ~,& (\ref{eqn:test_merge}.2) \\
  \frac{x'-x-\epsilon_f}{\Delta{t}} & > \frac{x'-x+\epsilon_f}{\Delta{t}} & \text{if } v_\bot <   \frac{x'-x-\epsilon_f}{\Delta{t}} \wedge v_\top \ge \frac{x'-x+\epsilon_f}{\Delta{t}}  ~,& (\ref{eqn:test_merge}.3)\\
  \frac{x'-x-\epsilon_f}{\Delta{t}} & > v_\top                  & \text{if } v_\bot <   \frac{x'-x-\epsilon_f}{\Delta{t}} \wedge v_\top <   \frac{x'-x+\epsilon_f}{\Delta{t}} ~, & (\ref{eqn:test_merge}.4) \\
  \mathrm{false} & & \mathrm{otherwise} .
  \end{array} \right.
\end{align}

Case \ref{eqn:test_merge}.2 is impossible: it implies that the curve set was already empty after insertion of the previous point.
That would already have been remedied by starting a new segment.
Case \ref{eqn:test_merge}.3 is impossible as all variables are strictly positive.
This allows us to rewrite equation \ref{eqn:test_merge} without expensive division operations:
Instead of the extremal slopes $v_\bot, v_\top$ we store
  the time $\Delta t_\bot, \Delta t_\top$ and
  value $x_\bot, x_\top$ of the previous extrema.
Then a new segment starts iff.
\begin{align}
\frac{x_\bot - x + \epsilon_f}{\Delta t_\bot} &> \frac{x' - x + \epsilon_f}{\Delta t} \,\bigwedge \nonumber \\
\frac{x_\top - x - \epsilon_f}{\Delta t_\top} &< \frac{x' - x - \epsilon_f}{\Delta t} \label{eqn:opt_check} \\
\Leftrightarrow (x_\bot - x + \epsilon_f) \Delta t &> (x' - x + \epsilon_f) \Delta t_\bot \,\bigwedge \nonumber \\
                (x_\top - x - \epsilon_f) \Delta t &< (x' - x - \epsilon_f) \Delta t_\top             \nonumber ~.
\end{align}

The update of the stored extrema can be modified in the same
way. Update $x_\bot \leftarrow x'$ and $t_\bot \leftarrow t_1$ when
\begin{align}
                & &                                         v_\bot &< \frac{x' - x - \epsilon_f}{\Delta t} \nonumber \\
\Leftrightarrow & & \frac{x_\bot - x' - \epsilon_f}{\Delta t_\bot} &< \frac{x' - x - \epsilon_f}{\Delta t} \label{eqn:opt_update} \\
\Leftrightarrow & &     (x_\bot - x' - \epsilon_f) \Delta t      &< (x' - x - \epsilon_f)       \Delta t_\bot ~.\nonumber
\end{align}
Analogous for $x_\top$.
Implementing eq.~\ref{eqn:opt_check} and \ref{eqn:opt_update} requires
  13 additions, 6 multiplications, 2 conditional moves and one branch.
Although the operation count is higher, expensive division and branch operations have been omitted.

\section*{Results and discussion}

A library implementing the \huri algorithm as described above
  is available under a GPL-3 open source license at \url{https://github.com/biosystemanalysis/hrtc}.
To compare our compression algorithm with the state of the art,
  we additionally integrated it into the \textsc{trajectory-ng} library.
The merged library is available at \url{https://github.com/biosystemanalysis/tng}.

\subsection*{\huri outperforms existing compression methods}

\textsc{Tng} comes with a benchmark application, used to compare compression algorithms, here.
The benchmark applies velocity verlet integration to simulate $512$ particles with a harmonic well potential
  $U(\Delta x) = \sin(\min(\|\Delta x\|, \frac{\pi}{2})^2$ and an arbitrary mass $2$, and timestep $2\cdot10^{-4}$ with arbitrary units (a.u.).
Initially, the particles are distributed randomly in a $15\times16\times17$ (a.u.) cuboid and equilibriated for $10^5$ steps.
After equilibration, this simulation is run for $10^7$ more timesteps to generate the benchmark trajectory.
The transient phase is omitted in order to avoid artifacts when applying different sub-sampling rates.
The resulting benchmark trajectory is then compressed using either \textsc{Tng}s native compression algorithm or our \huri compression.
We compare compression rates while varying the sub-sampling rate, the number of frames per block, and the maximal error $\epsilon$. 
The results are depicted in Table \ref{table:compression_rate}.
The remaining parameters are held constant during this paper:
  chunk size is set to 1024 support vectors,
  and block size to 2048 frames.

\huri outperforms the \textsc{Tng} compression in all cases,
  except when a very coarse sub-sampling rate ($\lesssim 1:1024$) is combined with high spatial resolution ($\epsilon \lesssim 0.001$).
These cases are practically irrelevant for two reasons:
  First, in our example $\epsilon = 0.0001$ corresponds to \SI{18}{b} of position information per dimension---almost
    equal to the \SI{24}{b} of a single precision float mantissa.
  And second, the position inaccuracy introduced by low temporal resolution far exceeds the error bound $0.0001$
    even for minuscule particle velocities.

Furthermore, \huri appears to outperform  all compression methods investigated by \textsc{Marais, et al.}\cite{Marais2012}
As neither their implementation nor their test data is public, we have to cautiously compare our benchmarks
  despite them running on different datasets.
The best compression rate \textsc{Marais'} algorithms achieves is $20.8$
  with 1:2 sub-sampling, a \SI{12}{b} quantization and \SI{0.014}{\angstrom} positional error.
This rate is overachieved by \huri already at the much coarser 1:32 sub-sampling
  with much smaller error bound $\epsilon=0.0001$---yielding a large buffer to compensate for the different datasets being used.
At a comparable error rate ($\epsilon = 0.01$) and sub-sampling (1:2) \huri achieves a compression ratio of 3419.

\newcommand{\kb}[1]{
\FPeval\ratio{120117548/#1}
\FPeval\round{round(\ratio:0)}
\FPeval\size{round(#1/1024:1)}
\SI{\size}{MiB} / 1:\SI{\round}{} }

\newcommand{\sz}[1]{
\FPeval\size{round(#1/1024:1)}
\SI{\size}{MiB}}

\newcommand{\rt}[1]{
\FPeval\ratio{120117548/#1}
\FPeval\round{round(\ratio:0)}
 \SI{\round}{} }

\newcommand{\rtb}[1]{
\FPeval\ratio{120117548/#1}
\FPeval\round{round(\ratio:0)}
 \textbf{\SI[detect-weight]{\round}{}} }

\begin{table}
\center
\caption{Comparison of compression ratio (uncompressed size / compressed size)
  for different error bounds $\epsilon$ and sub-sampling rates.
The size of the original, uncompressed \textsc{Tng} file was \SI{114}{GiB}.
The compressed file sizes vary between \SI{796}{KiB} and \SI{7.0}{GiB}.
Note that the sub-sampling factor is also included in the compression ratio.}
\scriptsize
\begin{tabular}{lrrrrrrrrrr}
\toprule

\multirow{2}{1.2cm}{Sampling rate} &
\multicolumn{2}{c}{$\epsilon=1$} &
\multicolumn{2}{c}{$\epsilon=0.1$} &
\multicolumn{2}{c}{$\epsilon=0.01$} &
\multicolumn{2}{c}{$\epsilon=0.001$} &
\multicolumn{2}{c}{$\epsilon=0.0001$} \\
\cmidrule(r){2-3}
\cmidrule(r){4-5}
\cmidrule(r){6-7}
\cmidrule(r){8-9}
\cmidrule(r){10-11}

& \huri       & \textsc{Tng}          & \huri       & \textsc{Tng}          & \huri       & \textsc{Tng}          & \huri       & \textsc{Tng}          & \huri        & \textsc{Tng} \\
\midrule

1:1 & \rt{25508} & \rt{3132196} & \rt{32960} & \rt{3139872} & \rt{47908} & \rt{3188944} & \rt{108388} & \rt{3610992} & \rt{291956} & \rt{7329736} \\
1:2 & \rt{13280} & \rt{1566364} & \rt{18260} & \rt{1572516} & \rt{35136} & \rt{1619592} & \rt{90604} & \rt{2005048} & \rt{256512} & \rt{5430348} \\
1:4 & \rt{7228} & \rt{783120} & \rt{11768} & \rt{788508} & \rt{28632} & \rt{833868} & \rt{79764} & \rt{1218152} & \rt{226132} & \rt{3173192} \\
1:8 & \rt{4228} & \rt{391824} & \rt{8904} & \rt{396808} & \rt{24492} & \rt{439968} & \rt{70188} & \rt{802236} & \rt{198080} & \rt{1824716} \\
1:16 & \rt{2848} & \rt{196172} & \rt{7352} & \rt{200932} & \rt{21460} & \rt{240560} & \rt{61924} & \rt{587932} & \rt{171904} & \rt{1032888} \\
1:32 & \rt{2212} & \rt{98024} & \rt{6520} & \rt{102604} & \rt{18892} & \rt{140604} & \rt{53828} & \rt{389760} & \rt{153696} & \rt{573460} \\
1:64 & \rt{1856} & \rt{49276} & \rt{5776} & \rt{53656} & \rt{16700} & \rt{90668} & \rt{47184} & \rt{223616} & \rt{142152} & \rt{316816} \\
1:128 & \rt{1636} & \rt{24896} & \rt{5016} & \rt{28972} & \rt{14644} & \rt{65132} & \rt{42472} & \rt{127112} & \rt{131272} & \rt{173628} \\
1:256 & \rt{1452} & \rt{12700} & \rt{4420} & \rt{16484} & \rt{12916} & \rt{45836} & \rt{39020} & \rt{71496} & \rt{91656} & \rt{99716} \\
1:512 & \rt{1212} & \rt{6248} & \rt{3716} & \rt{9828} & \rt{10900} & \rt{25252} & \rt{31688} & \rt{37744} & \rt{51088} & \rtb{50688} \\
1:1024 & \rt{932} & \rt{2960} & \rt{2908} & \rt{6064} & \rt{8632} & \rt{12924} & \rt{18764} & \rtb{18576} & \rt{25712} & \rtb{22848} \\
1:2048 & \rt{796} & \rt{1648} & \rt{2468} & \rt{4540} & \rt{6736} & \rt{7072} & \rt{11328} & \rtb{9508} & \rt{14388} & \rtb{12180} \\

\bottomrule
\end{tabular}
\label{table:compression_rate}
\end{table}

\subsection*{Compressing below one bit per sample}

\begin{figure}
\input{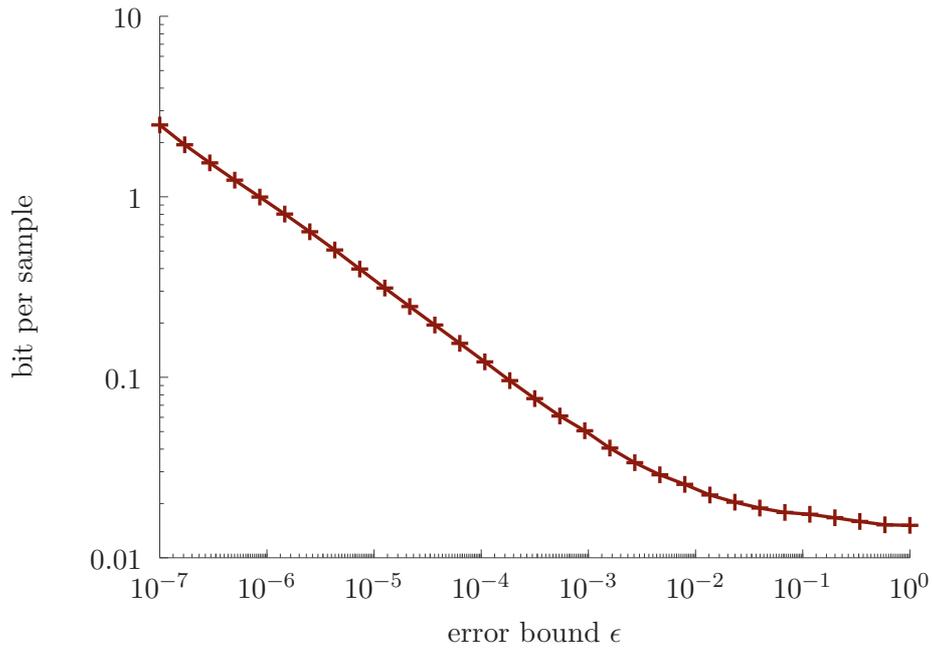}
\caption{\label{fluxkompensator}
  Average number of bits required to compress one sample---the
    position of one particle at one timepoint---depending on error bound $\epsilon$.
  A trajectory of $512$ particles over $10^7$ timesteps generated by the \textsc{Tng} benchmark application
    was compressed with 2048 frames per block.}
\end{figure}

To demonstrate the ability of our algorithm to reduce storage space needed per sample below the \SI{1}{b} limit,
  we compress a $2048$ frame $512$ particle simulation with varying error rates $\epsilon$.
For values of $\epsilon \tilde{>} 0.0000006$ the average space demanded per sample drops below $\SI{1}{b}$
  (see Figure~\ref{fluxkompensator}).

\subsection*{Compression is fast}

\begin{table}
\caption{Runtime (s) of an  MD simulation with the trajectory stored either with \textsc{Tng}, \huri, or not at all.
  The storage precision $\epsilon$ is varied, but has almost no effect on the runtime.}
\center
\begin{tabular}{rlllll}
\toprule
trajectory storage & $\epsilon = 0.0001$ & $\epsilon = 0.001$ & $\epsilon = 0.01$ & $\epsilon = 0.1$ & $\epsilon = 1$\\
\midrule
no storage & 1727.8 & 1730.1 & 1729.3 & 1727.0 & 1728.8\\
\huri compression & 1726.9 & 1727.2 & 1723.5 & 1728.4 & 1723.7\\
\textsc{Tng} compression & 1735.5 & 1723.5 & 1732.5 & 1729.9 & 1726.6\\ 
\bottomrule
\end{tabular}
\label{heartofgold}
\end{table}

We measured the throughput of compression and decompression with
  the same benchmark simulation used for the compression rate estimation.
As this simulation is executed on a general-purpose CPU,
  it requires relatively large amounts of time to compute the pairwise forces in each step.
This made the overhead of compression statistically insignificant compared to the run-time variations of the simulation itself (see Table \ref{heartofgold}).
To compare the throughput of different algorithms,
  we thus measured the time taken to compress a $\SI{114}{GiB}$ trajectory that was stored uncompressed on an SSD.

\begin{figure}
\input{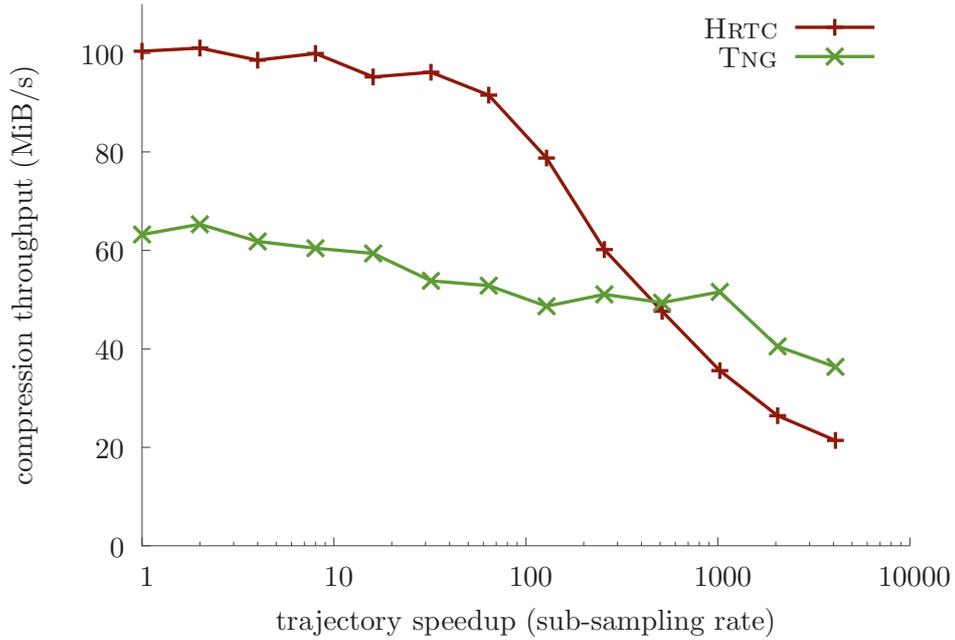}
\caption{\label{fig:speed}
  Compression throughput in \SI{}{MiB/s} of input data processed depending on trajectory velocity.
  Trajectories with faster movement were obtained by sub-sampling a common trajectory:
  A speedup of $n$ is equal to taking every $n$th frame.}
\end{figure}

The throughput of the tested algorithms depends on the magnitude of change of the trajectory to compress.
To simulate trajectories with different speed, we sub-sampled our test trajectory.
A 1:$n$ subsampling results in a $n$-fold speedup of the trajectory to compress.
The throughput of \textsc{Tng} and \huri compression are depicted in Figure~\ref{fig:speed}.
For slowly varying trajectories (speed-up below 100), \huri performs around $\SI{100}{MiB/s}$.
For speed-ups below factor $500$ \huri beats \textsc{Tng} in terms of throughput.
For fast rates of change, \huri's throughput converges against $\SI{20}{MiB/s}$.
On the same dataset, the general purpose \textsc{BZip2} compression achieves no more than $\SI{7}{MiB/s}$.

On their own data, \textsc{Marais} et al.~report compression rates between $13$ and $\SI{39}{MiB/s}$.\cite{Marais2012}

A second test with naked \huri compression---without the overhead of integration into the \textsc{Tng} library---reveals
  the extremal throughput:
A trajectory with constant particle position is compressed with $\SI{520}{MiB/s}$ and decompressed with $\SI{1838}{MiB/s}$.
Trajectories with purely random positions reach $\SI{65}{MiB/s}$ and $\SI{40}{MiB/s}$ for compression and decompression respectively.

All throughput tests have been performed on a single core of
  an Intel Xeon E5-2690 $\SI{2.9}{GHz}$ CPU.




\subsection*{Effect of error distribution between $\bm{\epsilon_q}$ and $\bm{\epsilon_f}$}

\begin{figure}
\input{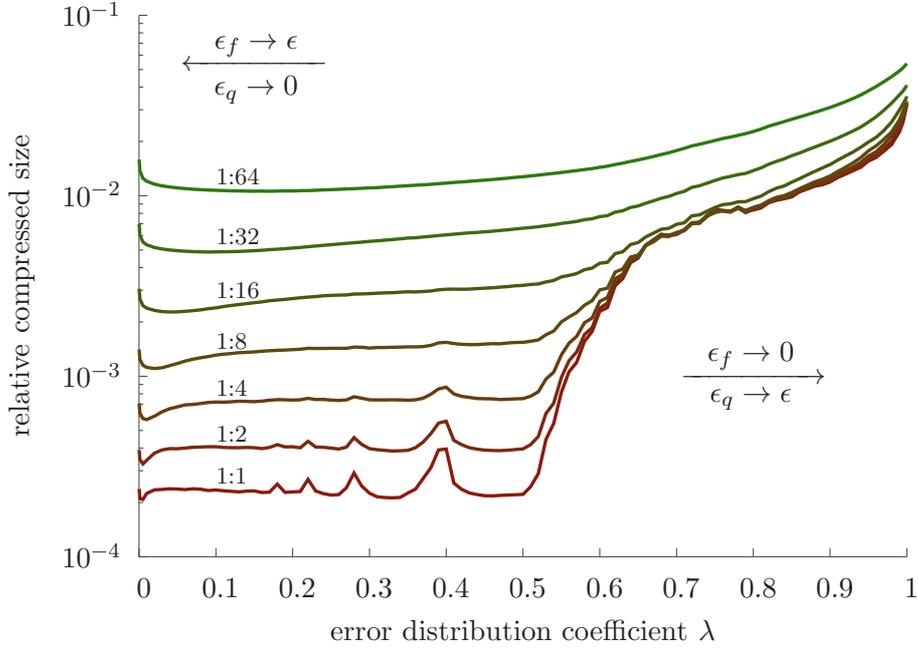}
\caption{\label{fig:qp-ratio}
  Compressed size depending on the error distribution coefficient $\lambda$ relative to the uncompressed file.
  Given the quantization error $\epsilon_q = \lambda \epsilon$ and the approximation error $\epsilon_f = (1 - \lambda) \epsilon$,
    for $\lambda \approx 0$ the algorithm degrades to a kind of linear regression, and
    for $\lambda \approx 1$ all error budget is used for quantization.
  The algorithm then degrades to a linear extrapolation scheme similar to the one used by \textsc{Marais et al}.\cite{Marais2012}
  The uncompressed data contained $10^4$ frames of $512$ particles with $3$ dimensions.
  It was sub-sampled at eight different rates (1:1 - 1:64).
  Compression was performed with total error $\epsilon = 0.01$, block size $10^4$ using plain \huri (without \textsc{Tng} metdata).}
\end{figure}

In the preceding sections we assumed the error budget $\epsilon$ is equally distributed
  between function approximation and quantization ($\epsilon_q = \epsilon_f = \frac{1}{2}\epsilon$).
The main motivation for that is to keep the number of parameters as small as possible.
To test whether this distribution is suitable, we parametrized the error distribution by $\lambda \in (0,1]$:
\begin{align}
\epsilon_q = \lambda \epsilon, \quad \epsilon_f = (1 - \lambda) \epsilon_q
\end{align}
We then compressed the \textsc{Tng} test data used above at different sub-sampling rates.
We compared the compression ratio achieved while varying $\lambda$.
The result is depicted in Figure~\ref{fig:qp-ratio}.
Between $0.01 \le \lambda \le 0.5$ a plateau of the compression ratio can be seen independent of the sub-sampling rate.
When spending more the $50\%$ of the error budget on quantization the compression rate sharply decreases.
This effect is the more pronounced, the more volatile the trajectory is (at sub-sampling rates below $1:16$).
We have observed similar behavior for all other datasets we tested (data not shown).

\section*{Conclusions}

We have developed a novel compression algorithm specifically for storage of molecular dynamic trajectories.
The algorithm is lossy, with a user-specified error bound.
By splitting the available error budget between quantisation error and function approximation error,
  we attain previously unachievable compression rates far below one bit per sample.

Even when saving with high fidelity (small time steps) compression rate and throughput are  outstandingly high.
Thus, we propose to use our format as primary representation of simulation data coming out of MD simulation kernels.
This will reduce the bandwidth demands between simulation kernel and analysis tools.
Using the computed linear functions as primal data representation
  will allow more integrated queries than the currently used uncompressed snapshots:
  for example a check for the minimal distance between two particles
    can be answered analytically on the level of segments, instead of iterating over all points.

Our approach offers some simple, yet rewarding extension points.
Foremost, parallelization of the compression is trivial:
  just compress subsets of the particles independently,
    preferably on the same cores that compute these subsets.
Due to better use of caches and smaller amounts of buffered support vectors
  when waiting for the termination of long-lasting function segments,
  we expect a super-linear speed-up from parallel execution.

Secondly, performance gains can be achieved by tight integration of function approximation and
  variable length integer representation (we used an external library so far).
This will save a whole pass over the data and allows to tune the integer compression parameters to the
  expected distribution occurring in our use-case.
Alternatively, an adaptive coder (e.g. entropy coder or arithmetic coder) could be employed.
While this would result in a modest improvement of the compression rate (we estimate at most factor 2),
  it would also reduce the throughput of our compression significantly.

Thirdly, integrating the \textsc{Tng} library and \huri more tightly should offer significant performance benefits.
We demonstrated that the throughput of our naked \huri library is much higher than
  our current integration into the \textsc{Tng} library.
There is an impedance mismatch between both libraries.
A deeper integration of \huri into \textsc{Tng}
  accompanied with some changes of \textsc{Tng}'s architecture
  to allow calling it inside the inner-most simulation loop without performance drawbacks
  will lift trajectory compression to a new level.

\subsection*{Author Contributions}

Jan Huwald designed the algorithm and implemented it in the \huri library.
Stephan Richter integrated \huri into \textsc{Tng}, generated the test data and executed the benchmarks.
The first draft was written jointly by Jan Huwald and Stephan Richter.
Peter Dittrich was responsible for project supervision and intensive revision of the draft.

\subsection*{Acknowledgments}

The authors acknowledge support from the European Union through funding under FP7–ICT–2011–8 project HIERATIC (316705).

\bibliography{paper}

\end{document}